\documentclass{sigchi}

\usepackage{balance}       
\usepackage{graphics}      
\usepackage[T1]{fontenc}   
\usepackage{txfonts}
\usepackage{mathptmx}
\usepackage[pdflang={en-US},pdftex]{hyperref}
\usepackage{color}
\usepackage{booktabs}
\usepackage{textcomp}

\usepackage{microtype}        
\usepackage{ccicons}          

\usepackage{todonotes}

\usepackage{pdfcomment}
\usepackage{comment}
\usepackage[autostyle]{csquotes}
\usepackage{paralist}
\usepackage{enumitem}
\setlist[enumerate]{itemsep=0mm} 

\def\plaintitle{Dungeons for Science: Mapping Belief Places and Spaces}

\def\emptyauthor{}
\def\plainkeywords{Belief space; cartography; role playing games, CSCW}

\makeatletter
\def\url@leostyle{%
  \@ifundefined{selectfont}{
    \def\UrlFont{\sf}
  }{
    \def\UrlFont{\small\bf\ttfamily}
  }}
\makeatother
\def\pprw{8.5in}
\def\pprh{11in}

\definecolor{linkColor}{RGB}{6,125,233}
\hypersetup{%
  pdftitle={\plaintitle},
  pdfauthor={\emptyauthor},
  pdfkeywords={\plainkeywords},
  pdfdisplaydoctitle=true, 
  bookmarksnumbered,
  pdfstartview={FitH},
  colorlinks,
  citecolor=black,
  filecolor=black,
  linkcolor=black,
  urlcolor=linkColor,
  breaklinks=true,
  hypertexnames=false
}

\begin{document}

\title{\plaintitle}

\numberofauthors{3}

\author{%
	\alignauthor{Aaron Dant\\
		\affaddr{ASRC Federal}\\
		\affaddr{Columbia, USA}\\
		\email{aaron.dant@asrcfederal}}\\
	\alignauthor{Philip Feldman\\
		\affaddr{University of Maryland, Baltimore County}\\
		\affaddr{Catonsville, USA}\\
		\email{phil@philfeldman.com}}\\
	\alignauthor{Wayne Lutters\\
		\affaddr{University of Maryland}\\
		\affaddr{College Park, USA}\\
		\email{lutters@umd.edu}}\\
}

\maketitle

\begin{abstract}
Tabletop fantasy role-playing games (TFRPGs) have existed in offline and online contexts for many decades, yet are rarely featured in scientific literature. This paper presents a case study where TFRPGs were used to generate and collect data for maps of belief environments using fiction co-created by multiple small groups of online tabletop gamers. The affordances of TFRPGs allowed us to collect repeatable, targeted data in online field conditions. These data not only included terms that allowed us to build our maps, but also to explore nuanced ethical problems from a situated, collaborative perspective.

\end{abstract}





\section{Introduction}

Games can function in social research as probes to reveal behavior, both individual and collective. They provide a bounded, manageable, \enquote{safe} environment to tease out beliefs and their ramifications. Considerable research has been done in this field, particularly with regard to \enquote {serious games}, where, for example, a surgical team can rehearse minimally-invasive surgery risk-free on a simulated patient \cite{willaert2012recent}, or users can learn to overcome their phobias \cite{botella2011treating}.

For group interactions, role-playing games (RPGs) provide considerable benefits as well. They have been shown to be effective in areas ranging from military training, to health care, and leisure.

Role-playing has three fundamental characteristics \cite{bowman2010functions}:
\begin{compactitem}
	\item it enhances a group's sense of communal cohesiveness by providing narrative enactment
	\item it encourages complex problem-solving and provides participants with the opportunity to learn an extensive array of skills through the enactment of scenarios
	\item it offers participants a safe space to enact alternate personas through a process known as identity alteration
\end{compactitem}

\subsection{Finding value in an overlooked genre}

Role-playing has been shown to be a valuable method for researching human behavior in combination with agent-based simulation \cite{guyot2006} however the forms of role-playing used tend to be specific to the subject matter. Only a few sociological studies exist which examine human behavior through the lens of tabletop fantasy role-playing games such as Dungeons \& Dragons \cite{fine2002shared}. Pat Harrigan sums this up nicely in the introduction to \textit {Second Person}: \enquote{in the last few years there has been much academic discussion of video games and other forms of digital media. but little that acknowledges in any depth the debt many of these forms owe to tabletop role-playing games. Further, it is not too much to say that where academic discussion of tabletop RPGs exists. it is largely cursory - and not infrequently wrong.} \cite{harrigan2010second}

From a CSCW perspective, TFRPGs have a compelling set of affordances for research, including:

\begin{compactitem}
	\item They present agreed upon and verifiable features
	\item They are a \enquote{subset} of reality, with fewer dimensions to explore, but with infinite choices within those dimensions
	\item Multiple groups can experience the same environment, making experiments easily repeatable
\end{compactitem}

This paper describes a process that uses the culture, structure and affordances of fantasy tabletop role-playing games (TFRPGs) to probe two related but distinct concepts:

The first is the \textit{belief place}, a cognitive structure shared between multiple groups about salient features in non-physical environments. In TFRPGs, these places could be rooms in a dungeon, or the country from which an imaginary character hails. 

The second, related concept is that of \textit{belief spaces}. These are the regions of choices associated with the place. What should I do in this room? Do I trust the ambassador from this imaginary kingdom? These are contextual choices that vary on the circumstance in the place at a particular time. These choices are constrained - their space is affected by the place that enables them.

The ability to portray and discriminate between the \enquote{ground truth} of belief places and the surrounding contextual belief spaces makes TFRPGs ideal for our research in creating maps that embody these concepts. In the next sections of this paper, we will walk through the background and culture of TFRPGs, the dungeon that was constructed to optimize our data gathering, some highlights from the data, and the resultant maps built from that data. We will conclude with a more general discussion about the utility of these special instruments.

\section{What is a TFRPG?}

The essence of role-playing is to take on the role of a character in an imaginary setting. This is something most of us have been doing in one form or another since we were children playing \enquote{make believe}. Role-playing has elements of improvisational acting, storytelling, and deductive reasoning. A player's character can be similar to their own personality, or embody everything that they are not.

tabletop fantasy role playing game, one player takes on the role of the Dungeon Master (DM or Game Master, Storyteller etc.) with the responsibility of describing the world the characters inhabit and telling the players what they can see, hear, feel, touch, and taste. The DM is also responsible for coming up with a story to set the players in. This story is not complete, but a framework that is collaboratively explored during the course of the game. The other players are the main characters in the story, and the DM adapts to the players' choices.

\subsection{Dungeons \& Dragons}
Dungeons \& Dragons (D\&D) is the first and best known fantasy tabletop role-playing game in the world. Open source copies of the basic rules are freely available \cite{dndbasicrules_2018}.  Players control an individual character who can cast spells or wield magic swords against fantastic, mythical creatures such as dragons and hobgoblins \cite{mackay2017fantasy}. Fellow wargamers Gary Gygax and Dave Arneson released D\&D to the public in 1977, where it has thrived in a wide variety of demographics and venues, ranging from tabletop play to online systems. The open source nature and broad availability of D\&D make it ideal for our research.

\subsection{Randomness: The Role of Rolls}

A core characteristic of D\&D is the inclusion of randomness in the form of dice rolls which allow chance to influence in the success or failure of player actions. A player character's skills may improve their chances of success at a particular action, but it is always possible to fail simply from chance.  Traditionally when played in person rolls are managed by a variety of different sided dice which allows for a finer granularity of influence.

\subsection{Characters}

In fantasy role-playing games there are a variety of powers afforded the players that give them beyond human abilities. These powers are used to confront challenges in the game, and represent a manifold of possible discussion spaces when navigating the environment. If each group has vastly different options with which to complete the objectives, the total number of participants will need to be expanded. However, some level of diversity needs to be included to provoke meaningful discussion. A party constructed of characters with identical abilities, will have less need for discussion or different perspectives. 

By the time of its first expansion \textit{Greyhawk} \cite{greyhawk1975}, Dungeons and Dragons had created four archetypal characters: 
\begin{compactitem}
	\item \textbf{Fighter}: offers direct combat strength and durability.
	\item \textbf{Thief} (Rogue): offers cunning and stealth.
	\item \textbf{Cleric}: provides support in both combat and magic.
	\item \textbf{Magic-User} (Mage): has a variety of magical powers
\end{compactitem}

To this day, official pre-generated characters for the \textit{Starter Set} are implementations of these same archetypes \cite{starter_set_characters_2014}. By using these well understood characters, we were able to limit the range of solutions for encounters to one manageable with a small number of participants. Groups were allowed to select multiple of the same archetype if desired, and to encourage an emotional connection with the characters, each player was allowed to assign the name, gender, and alignment for the pre-generated character they selected. Alignment is a D\&D convention for grounding a characters' moral beliefs. It is a two axis system that ranges from \enquote{lawful} to \enquote{chaotic} on one axis, and \enquote{evil} to \enquote{good} on the other, creating characters that can range from \enquote{chaotic evil} to \enquote{lawful good}\cite{dndbasic_1977} \cite{alignment1991}

\section{Case Study: The Trials of Tymora}

We chose TFRPGs because they are systems that take a latent structure (created by the DM), and, through the process of game-play, build a sequence of specific actions based on the interactions of the players and their evolving consensus of how to proceed through the current scenario \cite{borgstrom2005structure}. This provides a framework whereby multiple groups can come to consensus about many different kinds of problems, ranging from the practical (how do we get across the pit) to moral (do we kill the guard?). These vast narratives are capable of supporting intimate, honest discussion between the players, both in and out of character \cite{harrigan2010second}. The terrain that these narrative structures represent become Foucault's \textit{other spaces} \cite{foucault1986other}, and exist as alternate realities in the player's minds.

Why should we expect that small groups interacting in these spaces should produce the kind of data that we need to produce maps? The answer lies in the dimension-reducing process that groups achieve consensus. Moscovici showed that the process of consensus has several steps. First, the group has to implicitly agree on what they are going to discuss. All the individuals with their own, unique perspectives have to create a common, shared reality that they can then debate. Uncommon perspectives are literally incomprehensible on a neurological level to people who have not encountered them previously \cite{yeshurun2017same} \cite{parkinson2018similar}. Once the group has arrived at this lower dimensional space, they can either compromise and average the beliefs of the group, or achieve consensus around one of the poles of the discussion and move in that direction. This process of consensus moves the average belief of the group towards the pole of consensus \cite{moscovici1994conflict}. It is this process of dimension reduction and motion that makes us believe that it is possible to build low-dimension, human-comprehensible maps. To frame this discussion then, we must first describe maps, and how we might build them from co-created stories.

\subsection{What is a map?}
Maps are a special subset of diagram, that contain spatial and symbolic aspects such that it is possible to understand the geometric relations between symbolic elements \cite{fathulla2007diagram}. In a traditional map, an example of this difference would be a political map that shows country borders. The geographic (spatial) representations are present in the world, but the symbolic borders are a manifestation of beliefs that we as humans bring to the environment. 

Because we can \enquote{see ourselves in the map}, we can extrapolate from the places we have been to places that we have never seen. Maps can inform us of what we are likely to encounter on our journey. Building the maps \enquote{gives materiality and objectivity to space} \cite{jacob2006sovereign}. Finally, because of this shared representation that maps can convey at a glance, diagrams such as maps provide a type of communication that is distinct from other forms such as lists and stories. They afford reasoning about time and space \cite{isozaki1992mechanism}, even if the locations are imaginary \cite{rohl2008mapping}. 

\subsection{Mapping belief space}

Foucault describes the physical world as \enquote{the space of our primary perception} and imaginary worlds as \enquote{the space of our dreams and that of our passions}. Contrasting these worlds with idealized \enquote{utopias} he describes \enquote{these different spaces, these other places} as \textit{heterotopias} \cite{foucault1986other}. These are complex and related alternate realities. This view has also been espoused by Harris and Dourish who describe a similar relationship between the concept of space and place, while framing the debate in technological terms \cite{harrison1996re, dourish2006space}.

These \textit{belief spaces} may be very real to us. McNaughton et. al. and others have shown that there is a neural basis for our cognitive maps, and that these neural structures do not discriminate between real and imaginary spaces. \cite{mcnaughton2006path}

\subsection{Plotting belief trajectories}
Although \enquote{the map is not the territory} \cite{korzybski1958science}, one of the unique features of a map is the ability of readers to see themselves in it -- to recognize their surroundings. For a map to be useful as a map and not just a diagram, relationships between objects need to be portrayed in such a way that it is possible to \textit{plot a course} that accurately predicts what the traveler will encounter.  Lastly, a map needs to be able to support communication. One user should be able to communicate meaningful information about a course that another might be ready to embark on. As the goal of this research is to produce \textit{human-usable} maps, we structured the study with these qualitative measures in mind.

\subsection{Small Enough Stories}
Stories generate a shared understanding about the features that are encountered, the order in which they are encountered and the \enquote{physical} and cognitive elements of those features. In Lord of the Rings \cite{tolkien2012lord}, it is understood by readers that the Shire is to the West and that Mordor is to the East simply from the text. If Tolkien had not included a map of Middle Earth, this would still remain understood. Imagine if hundreds of stories were told throughout Middle Earth, each tracing a path across the features of that imaginary place. If we were to overlay those paths, and connect the matching features could we create a map simply from those shared adventures?

Middle Earth is vast and the number of stories required to cover that space would be immense, but within the context of TFRPGs we can design a smaller space which could be covered by fewer stories, and test our theory. The worlds created within these games are belief places where ground truth exists in the form of constructed maps and descriptive material prepared by the DM. These materials are used by the DM to describe and give context to the environment as the players explore it.

The larger the environment, and the more dimensions it contains the more agents and time are required to explore it fully. \cite{bellman1961curse} Narrative frames need to be constructed specifically to limit the number of dimensions with the goal of allowing a small group of players to conduct a thorough exploration of the environment.

By collecting these co-created stories of the participants in our online TFRPG, we show that it is possible to build maps that show the fictional environment overlayed with traces that indicate how the participants believed they should solve the problem of each room.

\section{Field experiment}
We designed our system as a field experiment to gather the data needed investigate the creation of maps of belief space. The affordances of TFRPGs are such that a ground truth is encoded in the structure of a D\&D adventure as a combination of maps and text prepared by the DM. We designed a world that afforded our navigational needs. Then we assembled and instrumented platforms to collect the data. We ran five full and one partial adventure and analyzed the results.

\subsection{Recruitment}

A variation of snowball sampling was used, where players were recruited from the DM and author's own online and physical player community. Additional members came from recommendations within this group. Each member was informed of the study and provided consent to participate in accordance with our human subject authorization.

\subsection{Player experience level}
The experience levels of the recruited players ranged from 30 years of experience to those who had only played D\&D once before. The range of knowledge the participants had from prior games provided us an opportunity to observe behaviors from different levels of experience as they interacted with the environment.

Experienced players of Dungeons \& Dragons have learned what is often referred to as \enquote{dungeon etiquette}. These learned behaviors include stealth, checking for traps frequently, and usage of adventuring tools (e.g., rope, pitons). This knowledge also includes tactical combat knowledge including never splitting the party up, sending the stealthiest character first to scout followed by a heavily armored character to take blows in combat, with the ranged and magic users following behind. Groups familiar with dungeon etiquette might be expected to easily bypass certain challenges which would be problematic for newer players.

Dungeons and Dragons is infamous for its \enquote{murder hobo} \cite{hutchings2016foucault} culture, where players are accustomed to using violence to solve conflicts, indiscriminately killing and looting their foes. Within the context of the fantasy game, it's frequently acceptable to murder \enquote{bad people} or really \textit{anybody} once they have engaged in hostilities. In the authors' experience (and in our study) the desire to consider negotiation or diplomacy is directly tied to how strong the players position appears to be, and their perceived risk. If they believe they can easily defeat the foes they will behave belligerently and aggressively, offering ultimatums or simply acting on their violence to get what they want. In situations where the party feels the outcome of violence is uncertain they carefully evaluate other opportunities. 

This inflection point is rare in Dungeons \& Dragons as the encounters are traditionally balanced around a \enquote{challenge rating} where the players encounter content which is suitable for their level. This was a constraint which required careful design consideration for our experiment as the goal was to provoke discussion. Our solution was to include a variety of challenge ratings ranging from the easily defeated to the impossible to defeat in combat with an \enquote{edge case} in between that is discussed in more detail in a later section.

\subsection{Playstyle}
Traditionally, tabletop D\&D takes place in co-located spaces with \textit{actual} tables. However, the long-standing tradition of  \enquote{play-by-post} (PBP) has been an text-based online equivalent of these tabletop environments for decades. Recently, chat-based versions of PBP systems have become increasingly popular. Because of the ability to recruit geographically far-flung and schedule-constrained groups and the ease of collecting textual data, we initially used traditional PBP. However, the exceedingly long playtime made maintaining user participation difficult. We were able to address this by moving to a chat-based system (Slack\footnote{\url{https://slack.com}}) which required longer contiguous periods of gameplay but shorter games overall. This is a hybrid model that maintains the realtime interaction of the tabletop, while affording online textual interaction.

\subsection{Platforms}
\begin{figure} [h]
	\centering
	\fbox{\includegraphics[width=1.0\columnwidth]{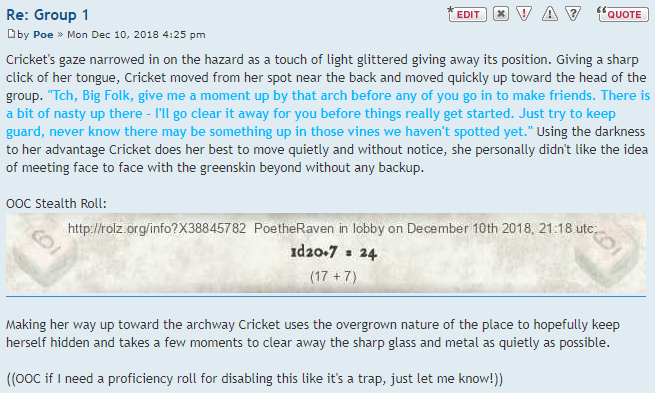}}
	\caption{BBS Post with dice roll}~\label{fig:PHPBB_post}
\end{figure}

Our initial PBP system was a PHP-based bulletin board system (BBS), instantiated and hosted on a private server\footnote{\url{https://antibubbles.com}}. Dice rolls were handled by rolz.org, a commonly used web application for verifiable dice mechanics for BBS-style environments. An example of a post with a dice roll is shown in figure \ref{fig:PHPBB_post}. Users were encouraged to use selected fonts and colors to differentiate between \textbf{dialog}, \textit{action}, and out-of-character (OOC) comments, though this information is not currently used in the analytics. We ran one full and one partial adventure with this system over the course of a month.

\begin{figure} [h]
	\centering
	\fbox{\includegraphics[width=1.0\columnwidth]{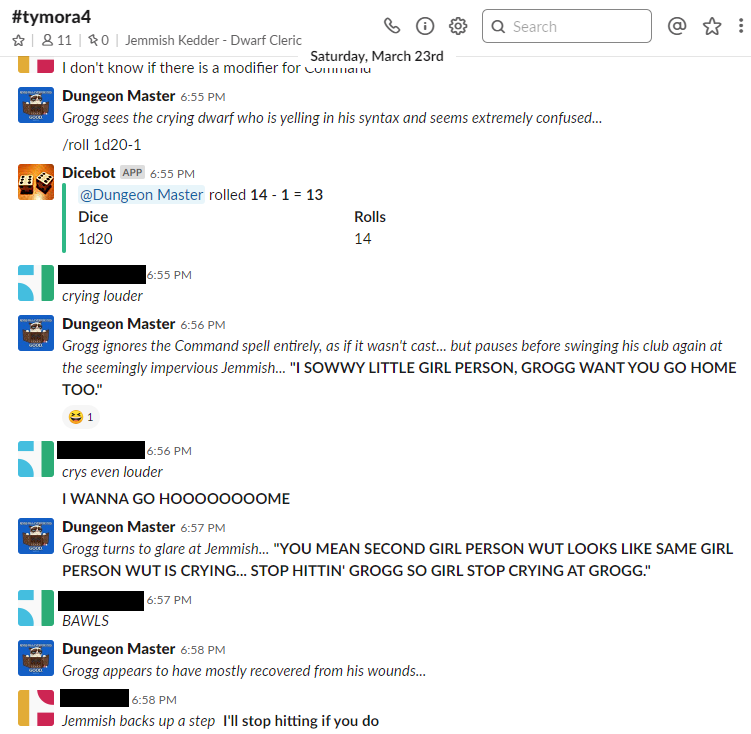}}
	\caption{Anonymized Slack chat with Dicebot}~\label{fig:slack_post}
\end{figure}

Setting up a Slack environment was simpler, only requiring the setup of the appropriate account and channels (one per run). A typical Slack session would cover two rooms and take approximately four hours. As such, a Slack adventure would take place over two sessions, though one run was a marathon nine-hour session. As with the BBS, the Slack environment had a \enquote{Dicebot} to manage dice rolls. As before, users were encouraged to use different font styles for dialog, action, and OOC comments. An example of the shorter, more rapid interactions using this system is shown in figure \ref{fig:slack_post}

\subsection{World building}

The game consists of a predefined play area (dungeon) with obstacles and challenges to overcome. The dungeon was created as linear and sequential, consisting of four rooms connected by magical gates that would close after the party members passed through them. This was determined to be complicated enough to develop the text analytics against, but also simple enough so that reasonable data could be gathered with relatively short runs. 

Once characters were established from the pre-generated set, they were run through the game, encountering the same scenarios in the same order. Positioning was an important factor for the first two rooms, so floorplans were provided along with the introductory text.  This pre-generated content information was provided to the players by the Dungeon Master, who pasted the content into the chat stream at the beginning of the adventure, the beginning of each room, and at the adventure's conclusion.

\subsection{Motivation and framing through backstory}
While a normal tabletop fantasy role-playing game has a general goal to create an enjoyable interactive narrative, the goal of this experiment was to create discussions about game features and opinions. The game consisted of a topologically simple linear dungeon with a variety of challenges. If each challenge only had a single solution, little conversation would be required by the participants, so special attention was paid to allowing multiple potential methods of success. Challenges were generated to incorporate stealth, diplomacy, violence, problem solving, and moral dilemmas, and framed in a context that would be familiar to most D\&D players:

\begin{displayquote}
	\textit{Near the port city of Waterdeep in Faerun, a strange new temple has appeared with the symbol of Tymora, goddess of luck and adventure. Word has spread throughout the region that adventurers visiting the temple will find a test set by the goddess herself and rewards beyond measure. Those emerging return with fantastic treasures, earning the honor and glory of besting the challenges within the Temple of Testing. Those who do not return... nobody knows what becomes of them.}
\end{displayquote}

Tabletop role-playing games are essentially co-created interactive fictional stories. They require a context to be enjoyable, and for the experiment to obtain meaningful results the participants had to perceive the experience as similar to other role-playing adventures. If the participants were thrown into a series of rooms without a larger story as context it would inhibit their suspension of disbelief and involvement with the environment. For the most recent version of Dungeons and Dragons (5th Edition) the default game world is \textit{Faerun}, otherwise known as \enquote{Forgotten Realms}. \cite{greenwood2001forgotten} This well documented and understood location gave a richer context to the game. With this in mind a plot to explain the experiment within the context of the game world was developed. 

The Temple of Trials is part of a wager between Tymora the goddess of luck and her dark sister Beshaba, the goddess of misfortune. The entire dungeon is an elaborate simulation designed to test real-life circumstances over and over with different populations to get statistically meaningful results to settle the wager. Beshaba contends that her sister and her followers have been incorrectly labeled as \enquote{Good} and her true nature is one of mercenary neutrality. While the players are unaware of the moral framing of the challenges, the cohesive story, and shared experiences over time gives each participant incentive to treat their decisions (and discussions) with real weight. Each player dedicated hours, or weeks of time to their character and this personal involvement seems to have dramatic impacts on the decision making process.

\subsection{Dungeon challenges}

Each group of participants encountered the same scenarios, and discussed different methods of solving the challenges. Describing the environment consistently was important to create a shared \textit{belief place} between individual runs. If the description had changed in important details between runs this would have introduced coherence problems where one group encountered and discussed features that did not exist for other groups. For two of the rooms where spatial movement was key to the challenge, top-down maps were created using Roll20.net to help illustrate the play area in more detail.

\subsubsection{Embedded tutorials}
\vspace{2mm}
\begin{displayquote}
	\textit{Ok, so. We can take on the goblins, or try to sneak down the stairs left of right after clearing the vines, is that correct? } Lorelei - Group 2
\end{displayquote}

The first room is intended to be an introduction, almost a tutorial to the experience of progressing through the dungeon. It has a clear objective (get through the golden gate at the far end), clear obstacles (the orc and goblin), and multiple methods of possible success. The orc and the goblin themselves are not particularly dangerous foes, so even a poorly executed plan will rarely result in a player death. Players become accustomed to exploring the environment by asking questions, performing actions, rolling dice, and discussing the situation with their companions. The level of detail this challenge uses helps to set the stage for the detail which can be expected throughout the gameplay.

Painstaking effort was taken incorporating small environmental details which both offered a sense of consistency and connection to the game world as well as a reassurance that this adventure would follow the patterns of traditional D\&D dungeons. The first challenge the players encounter in the starting antechamber is an archway overgrown with tangled dry vines. Moving through the vines may alert the likely hostile goblin and orc in the following room, and the grass underneath the archway has a large shattered chalice with sharp shards of glass and metal as well as nine silver coins. Their inclusion was with the hope that they would become recognizable features as each group interacted with these seemingly small details. The presence of the coins themselves were something of a red herring. Obtaining treasure is a huge part of the cultural experience of dungeons in role-playing games \cite{bowman2010functions}, and providing an early opportunity to demonstrate that \enquote{loot} and normal dungeon conventions were being followed.

One of the common features in dungeons is the presence of a variety of traps. These are generally mechanical or magical in nature and require problem solving, innovation, and lots of party discussion. The next room included deadly puzzles which need to be figured out as a group by the players, and required a different type of thinking and discussion than the goblin and orc encounter.

\subsubsection{Inducing discussion}
\vspace{2mm}
\begin{displayquote}
	\textit{lose the clanking, noisy, troll-waking armor} Imlodel
	
	\textit{what if the troll wakes up? I'll need my armor} Jemmish - Group 5
\end{displayquote}

The previous encounters can be dangerous, but it is unlikely the party will lose a player. This encounter introduces Grogg the troll, a non-player character (NPC) who is far more powerful than the game rules indicate is fair. Grogg is capable of killing a player character with a single blow, but he can be defeated, although only with great risk and luck. This encounter was designed to create discussion about what is the safest and best course of action. It was also created as a trap for players who intended to defeat the entire dungeon through superior use of combat tactics. Grogg offers the chance to win through brute force, but is unlikely to be defeated without at least one player death. 

\subsubsection{Conceptual problems}
\begin{figure}[h]
	\centering
	\includegraphics[width=0.9\columnwidth]{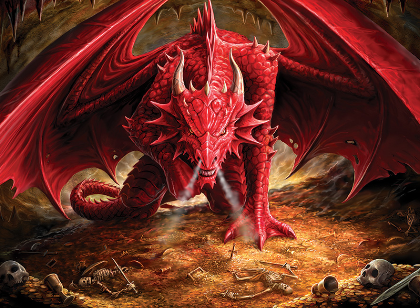}
	\caption{Ragosdias, the Red Dragon \protect\cite{annestokesdragon}}~\label{fig:dragon}
	
\end{figure}
\begin{displayquote}
	\textit{Someone has to leave in order to warn others. If the threat is real, only one has to die. If the threat is false, my choice means nothing} Gram - Group 4
\end{displayquote}

The last room was created to engage players with the traditional (one might even say ubiquitous) \enquote{trolley problem} \cite{thomson1976killing} to see how we could map a purely moral issue. In this room, the party discovers a massive treasure horde and with it, a deadly red dragon, one of the most fearsome and classic monsters in TFRPGs. It is immediately clear that there is no method of defeating the ancient beast through combat, and no visible escape from the room to use stealth or flight. Instead the dragon offers a game. The players must choose an old woman or a young boy to stay and starve, while the other will go free. If no choice is made, the dragon will devour the entire party and keep both the old woman and the boy in starving captivity. The problem is carefully worded to imply that the party must choose the boy or the woman, while also implying that there might be a possibility that one of the party can take their place.

\section{Results}

Five full and one partial adventures were run involving 23 participants over a period of several months. One complete session and one partial session were run using a PHP forum based system that stored data in a MySQL database which was downloaded for offline analysis. The other four sessions were run on Slack. These conversations were downloaded as JSON files and then parsed and loaded into a separate MySQL database. Database \textit{views} were developed so that the same query could work in both databases. Combined these contained 9,709 posts.

\subsection{Co-created stories}
We chose tabletop fantasy role-playing games for this experiment specifically because they are shared \textit{collaborative} narratives. This ended up providing benefits we could not have anticipated because they were emergent phenomena. 

\subsubsection{Unexpected salience}
\vspace{2mm}
\begin{displayquote}
	\textit{Can we taunt it to break its own neck coming down the slippery stairs? } Tamish - Group 5
\end{displayquote}

While details were included throughout the dungeon in the hope that salient features would emerge, some of these features ended up being far more relevant than expected. Directly outside the vine-choked archway entering the dungeon, steep winding stairs had been included, which as a point of adding flavor and distinctness were made slick with mud and moss. The expectation was that these slippery stairs would add a small obstacle, but the extent to which the stairs themselves became a challenge could not have been anticipated. Every single group encountered difficulties navigating the stairs, resulting in numerous player characters plummeting down the stairs suffering damage. It became so common that players discussing it afterwards dubbed the stairs \enquote{The Inclinerator} as an homage to its lethal effectiveness. As a result, the stairs themselves became one of the core features of this room.

\subsubsection{The five approaches to Grogg}
\vspace{2mm}
\begin{displayquote}
	\textit{are you all thinking diplomacy isn't even worth trying at this point? It didn't really help in the first room, but I have a feeling this place has been made to trick and surprise us, so I'm trying not to leave any option out.} Cricket - Group 1
\end{displayquote} 

The risk Grogg the troll posed if angered was epitomized by the \enquote{Red Mist Club} created by players who ended up dying after engaging Grogg in combat. In one encounter, a party of experienced gamers had prepared for the fight well, and nearly beaten Grogg in battle when he managed to land some hits of his own and turn the fight. The instant recalculation the players showed, by switching to diplomatic tactics and asking for a parley was exactly what the encounter was intended to do.

Many of the participant groups had to negotiate with Grogg, \textit{after} half of their party had been killed in battle with him. In each case, owing to his regenerative powers, Grogg was willing to forgive the assaults and the survivors ended on good terms with the giant green monster. In one case, upon awakening Grogg the party resorted to singing songs and diplomacy rather than test their combat prowess. A combination of incredibly detailed passages of the songs they were singing and astonishing success with rolls the party not only convinced Grogg to give them the key to the exit, but to join them as a companion for the rest of the dungeon.

One surprising result was the extent to which Grogg would become so well liked by the participants. A conscious choice was made to portray Grogg as simple-minded, but genuine, and open for negotiation. His speech patterns combined a child-like mind with an \enquote{\textbf{ALL CAPS YELLING}} style that made him oddly endearing. After each group concluded the participants became huge fans of Grogg, rooting for his success in future runs against other player groups with the hashtag \enquote{\#teamgrogg}. Even participants with characters that died to the troll agreed he was an excellent NPC.

\subsubsection{Insights from the dragon}
\vspace{2mm}
\begin{displayquote}
	\textit{\enquote{Get him home.  And deliver my cut of earnings to the people of Phandalin near Neverwinter, my home}  With this, before anyone can stop him, Edmund turns to the dragon \enquote{I make a counter offer.  In exchange for them} motions to the two caged people \enquote{I offer myself to take their place.  I will remain.  I will starve.  You will lose two peasants, and in return you will gain all that I have to offer.  Edmund of house DeVir of Neverwinter.  The last of a noble bloodline of the ruling class.}} Edmund - Group 2
\end{displayquote} 

The trolley problem has experienced a resurgence in recent years as it provides an ethical framework to understand how we should train/program autonomous systems such as self-driving cars. Awad, et. al. built a website that crowdsourced this problem, incorporating millions of responses from all over the world. Using this amassed data, they determined a hierarchy of who should die in a collision. At one end of the spectrum are pets, criminals, and the elderly who fare poorly. At the other end are children. At the top of the hierarchy are the vehicle's passengers, that are spared most often in these scenarios \cite{awad2018moral}. 

Our results are different. In each of the five completed adventures, the participants decided to self sacrifice in order to save both NPCs. In fact there were often arguments between the characters about who should stay. 

Why this result? We believe that it might be that the hours invested in the game opens a \textit{belief space} that is not accessed in an environment where one merely has to click on an image to decide if an imaginary character lives or dies. This dichotomy in results implies that the dungeon participants' belief space is not the one that the MIT study participants were embedded in when they responded. The belief space of the dungeon is one that is reachable through the co-creation of narratives over time. What spaces are adjacent to the former or the latter? How do we get from a space that seems to emphasize selfishness to one where the selfless act emerges repeatedly? Maps have the capacity to show us these relationships.

\section{Conclusions}

\textit{\enquote{By simplifying and exaggerating, games tell us what is real...} - G. Fine} \cite{fine2002shared}

In this paper, we have described a process that applies TFRPGs to produce data that can be used for a variety of research ends, including group dynamics, ethical deliberation, and a proof-of-concept cartography. We have demonstrated the effectiveness of this approach to leverage the unique benefits of TFRPGs in order to create maps of belief spaces and places. These were evaluated by the participants on their ability to represent their own travels and give opportunities to navigate the environment.

TFRPGs have several unique elements that make them powerful instruments for social science:
\begin{enumerate}
	\item By using fantasy as an abstraction they have the ability to address social and ethical issues from different perspectives \cite{bowman2010functions}
	\item They use small groups, building consensus in prolonged contact \cite{moscovici1994conflict}
	\item They are repeatable environments with ground truth
	\item They produce emergent patterns that are mappable, along with the trajectories taken through them \cite{feldman2018simon}
\end{enumerate}

In many ways, this meets the definitions of Foucault's heterotopias \cite{foucault1986other}, where the \enquote{function within society} exists outside of the particular run of a game. Rather it exists in the aggregation and results from the analytics. For scientific research the fact that these discussions are collaborative, emergent, and reproducible is a huge boon. The environment can be modified and leveraged to limit the range of options or create particular forms of discussion as we have shown. The depth of context available for TFRPGs specifically helps to create the player investment which produces these unique interactions.

Consider a fantastical case where the moral implications of magical weapons of mass destruction are debated by characters who have been fighting for their lives on the front lines of a desperate fictional war for survival. The participants have tremendous investment in both their characters, and the fictional land they are fighting to protect. The decision to authorize a first-strike can be debated within the context of powerful spells and the fictional kingdom rather than the immediate and reflexive response to a real-world discussion of nuclear weapons.

Research into similar complex social issues that requires participants' moral and emotional perspective may benefit greatly by the use of TFRPGs. Particularly as group interactions create clear textual signals that can be easily be identified in applications like mapping, as we have shown. 

Leveraging TFRPGs for science is similar to other forms of recruited social field experiments, however the downside is the pronounced social pressure against applying games to research. As Gary Fine noted, \enquote{Sociologists who study leisure typically find themselves attacked on two fronts. First they are accused of not being sufficiently serious about their scholarly pursuits. Second they are accused of alchemically transforming that which is inherently fascinating into something as dull as survey research computer tapes.} Our hope is that our successful case study contributes to the understanding of the value TFRPGs offer... that you can in fact make Dungeons for Science!

\section{Acknowledgments}
This effort would not have been possible without the players who spent hours killing, being killed (or nearly killed), singing to trolls and talking back to dragons in our dungeon:

{\small
\textbf{PHP Group 1}
\begin{compactitem}
\item William Ortiz - Enoch Isaac Jameson (Human Fighter)
\item Linda Gregory - Idril Sundancer (Elven Wizard)
\item Emily Smith - Cricket (Halfling Rogue)
\item Rachel Dant - Brin Daire (Human Fighter)
\end{compactitem}

\textbf{PHP Group 2}
\begin{compactitem}
\item Ian McBride - Reed Tosscobble (Halfling Rogue)
\item Ryan Smith - Edmund DeVir (Human Fighter)
\item Kristen Collins - Syllana (Elven Wizard)
\item Brandi Dant - Solgra Rubyhammer (Dwarven Cleric)
\end{compactitem}

\textbf{Slack Group 1}
\begin{compactitem}
\item Cynthia Flatley - Lorelei (Elven Wizard)
\item David Golden - Shelton Herrington (Halfling Rogue)
\item Michael McIver - Yenadar (Human Fighter)
\item Jim Hug - Thaldraed Ebonhand (Dwarven Cleric)
\end{compactitem}

\textbf{Slack Group 2}
\begin{compactitem}
\item Ryan Smith - Edmund DeVir (Noble Human Fighter)
\item Shelby Richardson - Ember (Elf Wizard)
\item RJ Walters - Merric Greenbottle (Halfling Rogue)
\item Bee Kuhlman - Esvele Greycastle (Human Fighter) 
\end{compactitem}

\textbf{Slack Group 3}
\begin{compactitem}
\item David Monath - Eod (Halfling Rogue)
\item Jenni Monath - Elra (Dwarven Cleric)
\item Bjorn Hasseler - Harbek Torunn (Dwarven Cleric)
\item Aaron Buchanan - Gram (Elven Wizard)
\end{compactitem}

\textbf{Slack Group 4}
\begin{compactitem}
\item Sherman Cater - Roscoe Tealeaf (Halfling Rogue)
\item Angel Cater - Tamish Kedder (Dwarven Cleric)
\item Debbie Wolf - Jemmish Kedder (Dwarven Cleric)
\item Mason Wolf - Imlodel Lightbringer (Elven Wizard)
\end{compactitem}
}
Extra thanks go to Emily Smith and Linda Gregory for participating not only as players, but also as Dungeon Masters for Slack Group 2 and Slack Group 3 respectively. Acknowledgment to Michael McIver for the many years of encounter building guidance that helped the author construct this dungeon.

%
%
%
%
%
\balance{}


\end{document}